\documentstyle[prl,aps,twocolumn]{revtex}
\begin{document}
\draft
\title{Comment on ``Electronic Properties of Armchair Carbon Nanotubes:
 Bosonization Approach''}
\maketitle

\narrowtext

In a recent Letter, Yoshioka and Odintsov (YO) \cite{od} 
have presented a bosonization approach aimed at a detailed
understanding of electronic properties of single-wall armchair
carbon nanotubes. Taking into account all possible 
Coulomb interaction processes among
the electrons, their central finding was that our previously published
field theory of nanotubes \cite{eg} is incomplete and 
should be supplemented by 
additional terms in the bosonized Hamiltonian. 
According to YO, these terms come about in the course of
a correct treatment
of the microscopic $C_3$ symmetry of the graphite lattice.
As we explain in the following, we disagree with this assertion.
The term omitted from our
presentation of the low-energy theory in Ref.\cite{eg} has in fact
been kept at intermediate stages of the calculation, and is
in any case automatically generated by the other terms under
the renormalization group (RG), without the need of going
back to the microscopic lattice description.
In addition, we show that at weak coupling the new term is irrelevant, and
at strong coupling it does not affect the low-energy physics.

Away from half-filling, backscattering gives rise to two contributions.
In the bosonized version, they read 
\begin{eqnarray*} 
H_1 &=& \frac{b}{(\pi a)^2} \int dx \; \{ \cos(2\theta_{\rho-})
\cos(2\theta_{\sigma-}) \\
&& \quad + \cos(2\theta_{\rho-}) \cos(2\phi_{\sigma-}) \} \;, \\    
H_2 &=& \frac{b'}{(\pi a)^2} \int dx  \, \cos(2\theta_{\sigma+})
\cos(2\phi_{\sigma-}) \;,
\end{eqnarray*}
where $b=[V_{pp}(2K_0)-V_{p-p}(2K_0)]/2$ and 
$b'=[V_{pp}(2K_0)+V_{p-p}(2K_0)]/2$ 
in the notation of YO.  Contrary to their statements,
the term $H_2$ has been kept in our RG
 analysis (it is exactly the term
$V_5$ in Sec.~4 of Ref.~\cite{eg2}), albeit
the RG equations have only been studied for an
initial value $b'=0$.  Let us now discuss that (i) the initial
values of the coupling constants 
given by YO are not trustworthy, and (ii) that $H_2$ is a 
marginally irrelevant operator that does not alter the low-energy
physics.

Concerning coupling constants \cite{momentum},
it is important to realize that detailed estimates 
are difficult to make within the bosonization approach  because
their values very sensitively depend on non-universal cutoffs. 
Changing the ratio of the interaction to the lattice 
cutoff, $a_0/a$, by only a factor of two changes the resulting
couplings $b$ and $b'$ by orders of magnitude.  

Remarkably, even if the estimates of Ref.~\cite{od} were correct, 
the RG equations \cite{eg2} show that at weak coupling,
the contribution $H_1$ is  the {\sl only}\,
(marginally) relevant operator.  Therefore it still makes
sense to first put all other coupling constants to zero (in particular
$b'$), as they all represent (marginally) {\sl irrelevant}\, operators,
and to study the RG flow due to $H_1$ alone \cite{eg}. 
Only at the new (intermediate) fixed point 
characterized by effectively large values of $b$, 
terms like $H_2$ or the nonlinear
forward scattering contribution \cite{eg} become relevant and then have
to be taken into account.  However,
as stated by YO themselves,  the resulting ground state is 
characterized by gaps in all modes except the total charge channel.
Therefore the term $H_2$ has
no physical effect and can safely be omitted in the low-energy
field-theoretical description of metallic carbon nanotubes,
particularly when the very small values of the relevant gaps
are taken into account.

After discussing these issues from a rather technical perspective, let us now 
give a simple physical argument.
As argued by YO, their corrections should be due to 
the correct $C_3$ symmetry of the lattice structure.
However, this argument overlooks a $U(1)$ symmetry 
emerging at low energy scales \cite{eg2}.
As is typical for critical field theories, 
terms breaking this symmetry down to the lower $C_3$ symmetry of the 
microscopic honeycomb lattice are irrelevant.  Hence the
lattice description pursued in Ref.~\cite{od} does not 
provide new information.
As a simple consequence of the $U(1)$ symmetry, 
the theory then applies to all metallic nanotubes and
not only to armchair tubes, in contrast to the assertion of YO.
As long as no gaps in the charged sector are present,
different wrapping indices will only affect the precise values
of the coupling constants but not the structure of the theory itself.

\vspace{0.5cm}

\noindent R.~Egger$^1$ and A.~O.~Gogolin$^2$
\vspace{0.2cm}

{\small ${}^1$ Fakult\"at f\"ur Physik

Albert-Ludwigs-Universit\"at

Hermann-Herder-Stra{\ss}e 3

D-79104 Freiburg, Germany }

\vspace{0.1cm}

{\small ${}^2$ Department of Mathematics

Imperial College

180 Queen's Gate

London SW7 2BZ, United Kingdom }


\begin{references}
\bibitem{momentum} We note that the correct momentum transfer 
is $K_0=2\pi/3a$ rather than YO's value $K_0=4\pi/3a$.
\end{references}
\end{document}